\documentclass{article}

\usepackage{arxiv}

\usepackage[utf8]{inputenc} 
\usepackage[T1]{fontenc}    
\usepackage{hyperref}       
\usepackage{url}            
\usepackage{booktabs}       
\usepackage{amsfonts}       
\usepackage{nicefrac}       
\usepackage{microtype}      
\usepackage{lipsum}		
\usepackage{amsmath}
\usepackage{graphicx}
\usepackage{natbib}
\usepackage{doi}
\usepackage{graphicx}
\usepackage{bm}
\usepackage{amssymb}
\usepackage{multicol}
\usepackage{multirow}
\usepackage{float}
\usepackage{enumitem}
\usepackage[capitalize]{cleveref}
\usepackage[title]{appendix}
\usepackage{tcolorbox}

\newcommand{\dbin}{\mbox{Binom}}
\newcommand{\dmulti}{\mbox{Multinom}}
\newcommand{\logit}{\mbox{logit}}
\newcommand{\Var}{\mbox{Var}}

\newcommand{\pmayin}{p^{mb|ni}}
\newcommand{\pmay}{p^{mb}}
\newcommand{\ps}{p^{c}}

\newcommand{\pins}{p^{i|c}}
\newcommand{\pks}{p_k^c}
\newcommand{\psmay}{p^{c|mb}}
\newcommand{\Mi}{M^{i}}
\newcommand{\My}{M^{yes}}
\newcommand{\Mn}{M^{no}}
\newcommand{\Mm}{M^{mb}}

\newcommand{\Mms}{M^{mb,c}}
\newcommand{\Hs}{H^{c}}
\newcommand{\Hi}{H^{i}}
\newcommand{\Ho}{\hat{H}_0}
\newcommand{\MB}{\mathcal{M}_{basic}}
\newcommand{\MI}{\mathcal{M}_{id}}
\newcommand{\MC}{\mathcal{M}_{class}}

\newtheorem{innercustomasm}{Assumption}
\newenvironment{customasm}[1]
  {\renewcommand\theinnercustomasm{#1}\innercustomasm}
  {\endinnercustomasm}

\title{Plant-Capture Methods for Estimating Population Size from Uncertain Plant Captures}

\date{} 					

\author{Yiran Wang \quad  \quad  \\
	Department of Statistics and Acturial Science\\
	University of Waterloo\\
	Waterloo, ON N2L 3G1 \\
	\And
	Martin Lysy \\
	Department of Statistics and Acturial Science\\
	University of Waterloo\\
	Waterloo, ON N2L 3G1 \\
	\AND
	Audrey B\'eliveau$^*$ \\
	Department of Statistics and Acturial Science\\
	University of Waterloo\\
	Waterloo, ON N2L 3G1 \\
        \texttt{audrey.beliveau@uwaterloo.ca}\\
}

\renewcommand{\shorttitle}{Plant-Capture Methods}

\begin{document}
\maketitle

\begin{abstract}
Plant-capture is a variant of classical capture-recapture methods used to estimate the size of a population. In this method, decoys referred to as ``plants'' are introduced into the population in order to estimate the capture probability. The method has shown considerable success in estimating population sizes from limited samples in many epidemiological, ecological, and demographic studies. However, previous plant-recapture studies have not systematically accounted for uncertainty in the capture status of each individual plant. In this work, we propose various approaches to formally incorporate uncertainty into the plant-capture model arising from (i) the capture status of plants and (ii) the heterogeneity between multiple survey sites. We present two inference methods and compare their performance in simulation studies. We then apply our methods to estimate the size of the homeless population in several US cities using the large-scale ``S-night'' study conducted by the US Census Bureau.
\end{abstract}


\section{Introduction} \label{intro}

Plant-capture \citep{laska1993plant} is a variation on Petersen capture-recapture experiments \citep{petersen1896yearly} employed to estimate the size of a target population. Unlike traditional capture-recapture experiments which utilize two distinct sampling occasions, the plant-capture method operates with just one capture occasion. This capture occasion is carried out on a population where marked individuals, referred to as ``plants", have already been introduced. Much like other capture-recapture sampling schemes, this method operates under the assumption that planted individuals cannot be distinguished from the rest of the individuals in the target population during sampling. A critical piece of information derived from this approach is the proportion of planted individuals that were successfully captured. This proportion provides insight into the capture probability, a fundamental factor required for estimating the size of the target population from an incomplete count. Under the plant-capture setting, \citet{laska1993plant} proposed a maximum likelihood estimator (equivalent to the Petersen estimator), a Chapman-Bailey estimator \citep{chapman1951some, bailey1952improvements}, and interval estimation for the target population size. Several papers have further discussed and extended this method \citep{martin1997issues, goudie1998plant, goudie2000conditionally, goudie2007plant, ashbridge2008conditionally}. 

Plant-capture methodology has been applied in different areas of survey research~\citep{ashbridge2008conditionally}, such as ecology \citep{skalski1982mark, yip1993estimating}, software reliability \citep{duran1981capture, yip1999sensitivity} and public health \citep{martin1992assessment,mccandless2016bayesian}. A common application of the plant-capture method is in estimating the size of homeless populations from point-in-time street surveys. As noted by \citet{berry2007repeated}, homeless surveys that omitted the inclusion of plants to adjust homeless counts have faced criticism for overlooking a substantial 40-70\%  of out-of-sight homeless individuals, which shows the importance of utilizing plants. Notably, the \citet{guide2008} has endorsed the plant-capture method as an innovative approach for estimating the size of a homeless population. The method has been applied to several surveys \citep[e.g.,][]{hopper2008estimating,mccandless2016bayesian}, including the large-scale Shelter and Street Night Enumeration (``S-night'') survey of the homeless conducted in five major US cities in 1990 by the United States Census Bureau \citep{hopper1991final,martin1992assessment,laska1993plant,martin1997issues}.

Capture-recapture methods (including plant-recapture methods) typically operate under the assumption that once an individual is captured by an enumerator, the latter can tell with certainty if the individual is marked or not by the presence or absence of a mark. However, in street surveys of the homeless, survey enumerators may be counting individuals from a distance, thus unable to determine whether these individuals are plants or not since planted individuals do not have distinguishable marks. For convenience, we name this setting ``capture without identification". In such situations, determining which plants were captured may rely on the plants' self-assessment of whether they were counted or not. While some plants may be able to assess this with certainty, other plants' assessments may be uncertain, as exemplified in the S-night survey, where plants had to answer ``yes", ``maybe" or ``no" to having been captured, via questionnaire. To the best of our knowledge, the only study addressing  uncertainty in plant assessment is \citet{martin1997issues}, which simply considers two extreme scenarios; one in which uncertain assessments (``maybes") are all presumed captured and another in which they are all presumed not captured. While these two extremes can be used for sensitivity analysis, they cannot be readily synthesized to provide a single population estimate, nor its commonly accompanying measures of statistical uncertainty (e.g., standard errors, confidence intervals).

In this paper, we address uncertainty in plant captures by introducing a rigorous modeling framework that explicitly incorporates capture status uncertainty. We do so by adopting a missing at random (MAR) assumption \citep{rubin1976inference}, which allows a formal account of the uncertainty inherent in the capture process. Our approach offers a flexible computational solution providing both frequentist and Bayesian population size estimators. Several simulation studies are conducted to validate the accuracy of our approach, establishing its ability to yield population estimates with the desired coverage probability. 

Section~\ref{model} presents three plant-capture models under various situations. In Section~\ref{method} we describe two inference approaches, frequentist and Bayesian. The performance of the proposed models and the differences between the two inference approaches are assessed through simulation studies in Section~\ref{sim}. In Section~\ref{application}, we apply our methodology to the S-Night plant-capture survey \citep{hopper1991final}. A discussion of our findings and directions for further work are presented in Section~\ref{discussion}.

\section{Methodology} \label{model}

Plant-capture surveys involve two types of individuals: plants and individuals from the target population. The size of the target population $H$ is unknown and is the target of inference, while the total number of plants, denoted by $M$, is known by design. In the context of capture without identification, the total number of plants can be partitioned into three observed quantities: $M = \My + \Mm + \Mn$, where $\My$ and $\Mn$ represent, respectively, the number of plants who are certain about having been captured/not captured, while $\Mm$ represents the number of plants who are uncertain (i.e. plants could not determine if they were captured or not). 

Other available data include the total number of captured individuals $Y$, collected by the enumerator during the survey. To describe $Y$, we introduce Assumption~\ref{assump1}.

\begin{customasm}{1}\label{assump1}
    Plants who self-assessed as ``yes'' were captured and plants who self-assessed as ``no'' were not captured.
\end{customasm}

Assumption~\ref{assump1} describes the certainty in the self-assessment, i.e. the plants' self-assessments accurately represent their capture status. Under this assumption, we can express $Y$ as the sum of three components: $Y = \My + \Mms + \Hs$, where $\Mms$ is the number of plants with uncertain assessment that were captured and $\Hs$ is the number of captured individuals from the target population. It is noteworthy that $\Mms$ and $\Hs$ are latent variables. Though we know the sum of $\Mms$ and $\Hs$ via $Y-\My$, we cannot acquire them individually because we cannot tell whether a captured individual is a plant or not solely by observation.

In order to leverage the information on capture probability acquired from plants to estimate the target population size, we introduce Assumption~\ref{assump2}:

\begin{customasm}{2}\label{assump2}
    The capture probability of plants and that of target individuals are homogeneous across individuals and equal.
\end{customasm}

Assumption~\ref{assump2} is a fundamental assumption for the plant-capture method, making it possible to apply the estimated capture probability for plants to the target population and derive a target population size estimate.

With the above framework in place, we now propose a basic model to relate the observed and latent variables to the quantity of interest $H$, along with two extensions to this model to account for uncertainty in capture status.

\subsection{Basic Model for Uncertain Plant Captures (Model $\MB$)} \label{model1}

Our basic model for estimating the population size $H$ is
\begin{gather}
    (\My, \Mm, \Mn)\mid M \sim \dmulti(M; \bm{\theta}_\text{I})\label{eq:IM}\\
    \Mms \mid \Mm \sim \dbin(\Mm, \psmay)\label{eq:IMm}\\ 
    \Hs \mid H \sim \dbin(H, \ps)\label{eq:IH}\\
    Y = \My + \Mms + \Hs.\label{eq:IY}
\end{gather}

Here $\bm{\theta}_\text{I}$ is a vector of probabilities for the plants' self-assessments as ``yes'', ``maybe'' and ``no'' respectively; $\ps$ represents the capture probability; $\psmay$ denotes the probability that a plant was captured given that this plant is uncertain about having been captured.

Since $\My$, $\Mm$ and $\Mn$ are known, $\bm{\theta}_\text{I}$ can be estimated. However, $\ps$, $\psmay$ and $H$ are not identifiable unless additional assumptions are made. Thus, we introduce a crucial assumption:

\begin{customasm}{3.I}\label{assumpI3}
    For plants, being captured is independent of self-assessing as ``maybe'', i.e., $\psmay =\ps$.
\end{customasm}

Assumption~\ref{assumpI3} indicates that plants' capture status are missing at random (MAR), which means that the missing capture status of plants can be fully accounted for by their self-assessment, and that there are no other unmeasured variables affecting their capture status~\citep{rubin1976inference}. This assumption enables us to use the information from the plants who are certain about their capture status to estimate the capture probability of the plants who are uncertain. Otherwise, $\ps$ and $\psmay$ are not identifiable. Although inherently untestable, our consideration of this assumption stems from its application as a simplifying assumption in other contexts, particularly in HIV surveillance studies \citep{vansteelandt2000regression,gustafson2023parameter}. In this specific scenario, the uncertainty surrounding specific HIV test results bears resemblance to the uncertainty inherent in plant capture.

As a consequence of the introduced assumptions, we can express
\begin{equation}\label{eq:thetaI}
    \bm{\theta}_\text{I} = [\ps(1-\pmay), \pmay, (1-\ps)(1-\pmay)]',
\end{equation}
where $\pmay$ represents the probability that a plant self-assessed as ``maybe", and is independent of being captured based on Assumption~\ref{assumpI3}. Now $\ps$, which is a critical parameter of the model, can be informed by the ratio of $\My$ to $\My+\Mn$. Further, we can combine Equations~\eqref{eq:IMm}, \eqref{eq:IH}, and~\eqref{eq:IY} into
\begin{equation}
    Y-\My \mid \Mm \sim \dbin(H + \Mm, \ps).\label{eq:Iui}
\end{equation}
Therefore, we can easily derive the posterior inference of $H$ from Equations~\eqref{eq:IM} and~\eqref{eq:Iui}, with $\bm{\theta}_\text{I}$ specified as in \eqref{eq:thetaI} . We henceforth denote this basic model as $\MB$.

\subsection{Incorporating Partial Identification Data (Model $\MI$)} \label{model2}

In Section \ref{intro}, the identity of captured individuals was unknown during the survey. However, in some plant-capture designs, it may be possible to acquire the identity of some captured individuals, as they may be identified via direct contact (e.g. interview) instead of simply being counted. In contrast to the ``capture without identification" design considered earlier, we name this design ``capture with (partial) identification". An ideal scenario would be that the identities of all captured individuals are known, where we can apply traditional plant-recapture methods to directly estimate the capture probabilities and subsequently derive population size estimates. But obtaining complete identification data in practice is often challenging. For example, in the S-Night survey described in Section~\ref{application}, enumerators were instructed to interview all individuals encountered in the survey sites who were not in uniform, engaged in money-making activities, sleeping or covered by sleeping bags or blankets between the hours of 2:00 and 4:00 a.m. \citep{martin1997issues}. Despite these instructions, only 15\% to 70\% of plants sighted were interviewed, depending on the city. Therefore, identification data may only be available for part of the augmented population. Still, incorporating this partial identification data can provide valuable insights into capture probabilities and enhance the accuracy of our population size estimates.

In the capture with partial identification design, we introduce a new parameter specifically for captured individuals, which is the probability of being identified, denoted by $\pins$. Furthermore, to facilitate the exchange of information between plants and the target population, we also introduce a new assumption:

\begin{customasm}{4}\label{assump4}
    All captured individuals have the same probability of being identified.
\end{customasm}

In this study design, there is certainty regarding the captured status of identified individuals, rendering self-assessment unnecessary for this subgroup. Consequently, the total number of plants can be decomposed into four observed quantities: $M = \Mi + \My + \Mm + \Mn$, where $\Mi$ is the number of identified plants while the definitions of $\My, \Mm$ and $\Mn$ are similar to those in model $\MB$ but pertaining to non-identified plants only. The captured count $Y$ can be partitioned into four components: $Y = \Mi + \My + \Mms + \Hs$ under Assumption~\ref{assump1}. In addition, we observe $\Hi$, the number of identified target individuals. A model for the ``capture with (partial) identification" framework is thus: 

\begin{gather}
    (\Mi, \My, \Mm, \Mn) \mid M \sim \dmulti(M; \bm{\theta}_\text{II})\label{eq:IIM}\\
    \Mms \mid \Mm \sim \dbin(\Mm, p^{c|mb,ni})\\
    \Hs \mid H \sim \dbin(H, \ps) \label{eq:IIH}\\
    \Hi \mid \Hs \sim \dbin(\Hs, \pins) \label{eq:IIHi}\\
    Y = \Mi + \My + \Mms + \Hs, \label{eq:IIY}
\end{gather}
where $p^{c|mb,ni}$ is the probability that a plant was captured given that it was not identified and that it self-assessed as ``maybe''.

With the identification data included in the model, we adapt Assumption~\ref{assumpI3} as follows:

\begin{customasm}{3.II}\label{assumpII3}
Among the non-identified plants, capture by an enumerator is independent of self-assessing as ``maybe'', i.e., 
    \begin{equation}\label{eq:MARII}
        p^{c|mb,ni}=\frac{\ps(1-\pins)}{\ps(1-\pins)+(1-\ps)},
    \end{equation}
where the right-hand side is the capture probability among the non-identified plants.
\end{customasm}

Assumption~\ref{assumpII3} limits the independence to the non-identified individuals, which means that the MAR mechanism is only assumed among the non-identified individuals since it applies to the plants' self-assessments.

As a consequence of the assumptions, we have
\begin{equation}\label{eq:thetaII}
\bm{\theta}_\text{II} =
    \begin{bmatrix}
        \ps\pins \\
        \ps(1-\pins)(1-\pmayin) \\
        \ps(1-\pins)\pmayin+(1-\ps)\pmayin \\
        (1-\ps)(1-\pmayin)
    \end{bmatrix},
\end{equation}
where $\pmayin$ represents the probability that a plant self-assessed as ``maybe'' given not identified. We can then estimate $H$, the size of the target population, by fitting the model described in Equations \eqref{eq:IIM} to \eqref{eq:IIY} but parameterized by $H$, $\ps$, $\pins$ and $\pmayin$ via Equations~\eqref{eq:MARII} and~\eqref{eq:thetaII}. We henceforth refer to this model as $\MI$.

\subsection{Incorporating Heterogeneity Between Sites (Model $\MC$)}\label{model3}

An important assumption for the plant-capture method is that capture probability is constant and equal for plants and individuals from the target population \citep{laska1993plant}. However, in practice, this assumption may be violated for various reasons -- especially when there are multiple sites enumerated in a survey. For example, in point-in-time street surveys of the homeless, there can be variations in the capture probability across different sites due to several factors such as visual barriers, drug activities, time constraints, or enumerator behavior. Suppose we classify the sites as ``hard'' if more than 50 percent of plants at one site mentioned any of these problems, and ``easy'' otherwise. It is reasonable to expect that the probability of being captured by an enumerator would be larger in ``easy" sites compared to ``hard" sites.

To account for such heterogeneity between sites, it is possible to introduce classes in our model. We assume that there are $K$ classes, and that capture probability varies across these classes, with $\pks$ denoting the capture probability of each individual in class $k\in\{1,...,K\}$. To formulate the new model, we define the following within each class $k\in\{1,...,K\}$:
\begin{gather}
    \nonumber(\Mi_k, \My_k, \Mm_k, \Mn_k) \mid M_k \sim \dmulti(M_k; \bm{\theta}_\text{III,k})\\
    \nonumber\Mms_k \mid \Mm_k \sim \dbin(\Mm_k, p^{c|mb,ni}_k)\\
    \nonumber\Hs_k \sim \dbin(H_k, \pks)\\
    \nonumber\Hi_k \sim \dbin(\Hs_k, \pins)\\
    Y_k = \Mi_k + \My_k + \Mms_k + \Hs_k \label{eq:IIIY}.
\end{gather}
Here, all notations have similar interpretations as in $\MI$ but are within class $k$. Note that $p^{c|mb,ni}_k$ is indexed by $k$ because the capture probability varies by class. While here we assume that $\pins$ and $\pmayin$ are homogeneous for the sake of demonstration, the model can also be generalized further to accommodate variations in these parameters across different classes by introducing additional parameters specific to each class.

For this class-based model, the assumptions defined for $\MI$ are applied within each class, notably the MAR structure is applied within each class (Assumption~\ref{assumpIII3}):

\begin{customasm}{3.III}\label{assumpIII3}
     Among the non-identified plants, capture by an enumerator is independent of self-assessing as ``maybe'' within each class, i.e., $p^{c|mb,ni}_k = \frac{\pks(1-\pins)}{\pks(1-\pins)+(1-\pks)},$ for $k=1,...,K$,
\end{customasm}

\noindent and
\begin{equation*}
\bm{\theta}_\text{III,k} =
    \begin{bmatrix}
        \pks\pins \\
        \pks(1-\pins)(1-\pmayin) \\
        \pks(1-\pins)\pmayin+(1-\pks)\pmayin \\
        (1-\pks)(1-\pmayin)
    \end{bmatrix}.
\end{equation*}

Hence, we can estimate the size of the homeless population $H_k$ within each class, and sum them up to obtain an estimate of the total size of the homeless population $H = \sum_{k=1}^K H_k$. We denote the class-based model described in this section by $\MC$.

\section{Inference Approaches} \label{method}

\subsection{Frequentist Inference via Maximum Likelihood} \label{mle}

Frequentist inference --  specifically maximum likelihood (ML) estimation -- is straightforward to conduct for model $\MB$. Based on Equations~\eqref{eq:IM} and~\eqref{eq:Iui}, the joint likelihood of the parameters of interest $\bm{\gamma}$ = ($H$, $\ps$, $\pmay$) is
\begin{align*}
    L(\bm{\gamma}; y, m^{yes}, m^{mb}, m^{no} ) &= P_{\bm{\gamma}}(Y=y,~\My=m^{yes},~\Mm=m^{mb},~\Mn=m^{no})\\
    &= \frac{(m^{yes}+m^{mb}+m^{no})!}{m^{yes}!m^{mb}!m^{no}!}\{\ps(1-\pmay)\}^{m^{yes}}(\pmay)^{m^{mb}}\{(1-\ps)(1-\pmay)\}^{m^{no}} \times \\
    & \left( \begin{array}{c} H+m^{mb} \\ y-m^{yes} \end{array} \right)(\ps)^{y-m^{yes}}(1-\ps)^{H+m^{mb}-y+m^{yes}}.
\end{align*}
Maximizing this joint likelihood with respect to $\bm{\gamma}$ yields the following ML estimators (MLEs): $\widehat{\ps} = \My/(\My+\Mn)$, $\widehat{\pmay} = \Mm/M$ and $\widehat{H} = \lfloor Y/\widehat{\ps} - M \rfloor$, where $\lfloor x\rfloor$ denotes the floor function. Derivations can be found in the Appendix.

In contrast, models $\MI$ and $\MC$ involve latent variables, such that the MLE of $H$ cannot be expressed in closed form. To address this challenge, we adopt a practical strategy of marginalizing out the latent variables numerically, and identifying the mode of the joint likelihood using numerical optimization techniques to obtain the MLE of each parameter. Additionally, we approximate the variance-covariance matrix using the inverse of the negative Hessian matrix of the log-likelihood evaluated at the ML estimates. By leveraging the asymptotic normality property of MLEs, we further provide confidence intervals for the estimated parameters. 

The probability functions for model $\MI$ and model $\MC$ can be expressed in a general form $L(\bm{\gamma}; \bm{x}) = P_{\bm{\gamma}}(\bm{X} = \bm{x}) = \sum_{\bm{z} \in \Omega} P_{\bm{\gamma}}(\bm{X}=\bm{x},\bm{Z}=\bm{z})$, where $\bm{\gamma}$ represents the model parameters, $\bm{X}$ denotes the observed data, $\bm{Z}$ represents latent variables, and $\Omega$ is the set of possible values for $\bm{Z}$. In model $\MI$, we have $\bm{Z} = \Mms$, because knowing $\Mms$ also provides information about the other latent variable $\Hs$ through Equation~\eqref{eq:IIY}. In this context, $\bm{\gamma} = (\log(H),\logit(\ps),\logit(\pins),\logit(\pmayin))$ and $\bm{X} = (Y,\Mi,\My,\Mm,\Mn,\Hi)$. By summing over $\bm{Z}$, we derive a marginal likelihood of $\bm{X}$ that excludes latent variables, thereby facilitating the direct application of the MLE method. Note that some parameters within the model represent counts, taking on positive integer values, while others represent probabilities that range between 0 and 1. Therefore, we apply a log transformation on the counts and a logit transformation on the probabilities, eliminating any constraints on their bounds to prevent computational issues arising from boundary constraints.

Determining $\Omega$ depends on the domain of $\Mms$ and the domain of $\Hs$. For $\MI$, we establish the bounds based on two constraints. The first constraint,
\begin{equation}
0\leq\Mms\leq\Mm,\label{eq:const1}
\end{equation} 
is simply due to the domain of $\Mms$. In addition, the relationship between $\Mms$ and $\Hs$ in Equation~\eqref{eq:IIY} implies $\Mms = Y - \Mi - \My - \Hs$, with $\Hs$ bounded within its domain $\Hi\leq\Hs\leq H$. The second constraint arises as a consequence:
\begin{equation}
    Y - \Mi -\My - H \leq \Mms \leq Y - \Mi - \My - \Hi.\label{eq:const2}
\end{equation}

\noindent Combining constraints \eqref{eq:const1} and \eqref{eq:const2}, we have $\Omega = [a_\text{II},b_\text{II}]$ with
\begin{align*}
    a_\text{II} & = \max(0, Y - \Mi - \My - H)\\
    b_\text{II} & = \min(\Mm, Y - \Mi - \My - \Hi).
\end{align*}

Similarly, for $\MC$ we use $\mathbf Z =(\Mms_1,\dots,\Mms_K)'$ and $\Omega = \Omega_1 \times \dots \times \Omega_K$ where $\Omega_k = [a_{k, \text{III}}, b_{k, \text{III}}]$ with
\begin{align*}
    a_{k, \text{III}} & = \max(0, Y_k - \Mi_k - \My_k - H_k)\\
    b_{k,\text{III}} & = \min(\Mm_k, Y_k - \Mi_k - \My_k - \Hi_k).
\end{align*}

Consider the log-likelihood $\log \{L(\bm{\gamma}, \bm{x})\}$ where $\bm{x}$ is the observed data from $\bm{X}$. The ML estimator of the vector of parameters $\bm{\gamma}$ is the mode of the log-likelihood which can be approximated numerically using the Nelder–Mead optimization method \citep{nelder1965simplex}. This method is easy to implement with the function \textit{optim} in R \citep{r}. To estimate the variance-covariance matrix, we numerically approximate the Hessian matrix using Richardson extrapolation, which can be carried out with the function \textit{hessian} from the package \textit{numDeriv} \citep{numDeriv} in R. The same numerical method is also applied to Model $\MB$ in this study for simplicity.

Notably, given that our models involve parameter transformations within $\bm{\gamma}$, it is essential to revert the MLEs ($\hat{\gamma}$), variance estimators ($\widehat{\Var(\hat{\gamma}})$), and confidence intervals (CIs) to their original scales. For MLEs, this is achieved by applying the corresponding inverse functions to the estimates. For the variance estimators, we use the delta method to transform them to the original scale. For the CIs, we first construct the 95\% CI of each parameter on the transformed scale as $\hat{\gamma} \pm 1.96 \sqrt{\widehat{\Var(\hat{\gamma}})}$. Then we apply the corresponding inverse functions on these CIs to obtain the CIs on the original scale.

\subsection{Bayesian Inference via MCMC} \label{bayes}

When dealing with complex models, marginalizing out all latent variables may prove inefficient or impractical. Given the hierarchical structure inherent in our proposed models, a Bayesian framework, coupled with Markov Chain Monte Carlo (MCMC) methods, emerges as an effective alternative for conducting inference. Notably, probabilistic programming languages for MCMC sampling have a broad appeal with applied scientists as they allow symbolic coding of a hierarchical model along with its priors. This allows users flexibility in extending or customizing models (e.g. sharing parameters across years or specifying exchangeable parameters via hierarchical priors). Additionally, these languages conveniently come equipped with built-in MCMC algorithms for posterior distribution sampling.

Among probabilistic programming languages for MCMC sampling, those based on BUGS \citep{gilks1994language} stand out for their unique ability to handle discrete latent variables, which is a feature present in our models. These languages include, for example, JAGS \citep{plummer2003jags} and NIMBLE \citep{nimble}. When using BUGS-based languages, we need to be careful about their grammar rules. For example, in Equations~\eqref{eq:IIY} and \eqref{eq:IIIY}, $Y$ is represented as the sum of two observed variables and two unobserved variables, while we have the observed value for $Y$ as data. In this case, one should use the \textit{dsum} function in JAGS so that the sampler will update the unobserved variables together ensuring that the sum constraint is preserved. There are also alternative ways, such as defining custom functions and distributions in NIMBLE or employing the zeros-ones trick \citep{ntzoufras2011bayesian}. The code to implement our models is provided in the Supplementary Materials.

Despite the simplicity of conducting Bayesian inference, one of the drawbacks of Bayesian inference is that it may be sensitive to the choice of the prior. Though we strive for noninformative priors when there is no prior knowledge available, the results could still be affected by the noninformative priors we choose. Further discussion on the choices of prior can be found in \citet{bda}, and specific choices of priors for our analysis will be presented in Section~\ref{sim}. Besides, when dealing with complex models, BUGS-based languages can be computationally expensive. To address this issue, refer to Appendix B for two alternative computational approaches that enhance efficiency, and Supplementary Tables in Appendix C for additional simulation study results.

\section{Simulation Study} \label{sim}

To evaluate the performance of our models, we conducted a simulation study for each of the models, leveraging information from the S-night survey to emulate real-world conditions in homeless population size estimation. In each study, we considered two distinct scenarios: small city and large city. In the context of the small city scenario, we set the true value of $(M,H)$ to (15, 150) for $\MB$ and $\MI$, and (30, 300) for $\MC$. For the large city scenario, the true value of $(M,H)$ is set to (100, 1,500) for all three models. This contrast in city sizes allows us to discern potential variations in the performance of our methods when applied to cities with varying homeless population sizes.

Under each of these six specified settings, we simulated 1,000 datasets. We set the true values for $(\ps, \pmay)$ to (0.7, 0.2) in $\MB$, and we set $(\ps,\pmayin,\pins)$ to (0.7, 0.2, 0.8) in $\MI$. In $\MC$, we assumed that there are two classes for the sites: easy and hard. For this setup, the true capture probability for easy sites was set to $\ps_1=0.9$ while the capture probability for hard sites was set to $\ps_2=0.4$. We also assumed that 60\% of the sites were easy while the remaining 40\% were hard, leading to an overall capture probability of 0.7, consistent with our setting for $\ps=0.7$ in the other studies. Furthermore, we distributed the plants into these site classes proportionally, with 60\% of plants in easy sites and 40\% in hard sites. The true values for $(\pmayin,\pins)$ were also set to (0.2, 0.8). 

The inference for each dataset was conducted both via Maximum Likelihood and Bayesian estimation, which we aim to compare. When performing Bayesian inference, we used independent Uniform$(0,1)$ priors on all parameters representing probabilities. For models $\MB$ and $\MI$, a log-normal prior (rounded to the nearest integer) with mean 0 and variance 100 was specified on the population size $H$; for model $\MC$ equivalent independent prior were defined for every $H_k$. The MCMC algorithm was implemented in JAGS, employing 3 chains, each comprising 30,000 iterations. We treated the first 15,000 iterations as burn-in to guarantee the stability and convergence of our results.

In each of our simulation studies, we employed a comprehensive set of evaluation metrics to assess the performance of Bayesian and MLE approaches. These metrics include: average estimates, average standard deviations (SD), relative Monte Carlo biases (RBias), relative root mean squared errors (RRMSE), coverage probabilities (CP) and average lengths of the 95\% confidence interval or credible interval (LCI). Note that for the Bayesian method, we used the expected posterior medians as estimates.

We begin with the simulation and analysis setting for model $\MB$, the results of which are shown in Table~\ref{tab:mod1}. In the small city scenario, the MLE of $H$ has a negligible bias, whereas the Bayesian estimator has a relative bias of 6\%. However, the coverage probability of the MLE is too low at 85\%. In the large city scenario, both estimators perform similarly with negligible biases and coverage probabilities close to 95\%. Additionally, we observe that although the two estimators have the same relative root mean squared errors, the Bayesian estimator yields a larger LCI compared to the MLE.

\begin{table}
    \centering
    \caption{Results of the simulation studies with $\MB$ for the MLEs and the Bayesian estimators. All the values are rounded to integers or 2 decimal points.}
    \label{tab:mod1}
    \resizebox{\textwidth}{!}{
    \begin{tabular}{ccccccccccc}
        \hline
        Method & $M$ & Parameter & True Value & Estimate & SD & RBias & RRMSE & CP & LCI\\
        \hline
        \multirow{3}{*}{MLE} & \multirow{3}{*}{15} & $H$ & 150 & 149 & 31 & -0.01 & 0.24 & 0.85 & 126\\
        & & $\ps$ & 0.7 & 0.73 & 0.12 & 0.04 & 0.19 & 0.98 & 0.48\\
        & & $\pmay$ & 0.2 & 0.20 & 0.10 & 0.01 & 0.51 & 0.98 & 0.42\\
        \hline
        \multirow{3}{*}{Bayesian} & \multirow{3}{*}{15} & $H$ & 150 & 159 & 43 & 0.06 & 0.24 & 0.97 & 160\\
        & & $\ps$ & 0.7 & 0.68 & 0.12 & -0.03 & 0.16 & 0.97 & 0.45\\
        & & $\pmay$ & 0.2 & 0.23 & 0.10 & 0.14 & 0.49 & 0.98 & 0.37\\
        \hline
        \multirow{3}{*}{MLE} & \multirow{3}{*}{100} & $H$ & 1,500 & 1,497 & 114 & -0.00 & 0.08 & 0.93 & 449\\
        & & $\ps$ & 0.7 & 0.70 & 0.05 & 0.01 & 0.07 & 0.95 & 0.20\\
        & & $\pmay$ & 0.2 & 0.20 & 0.04 & 0.00 & 0.20 & 0.94 & 0.16\\
        \hline
        \multirow{3}{*}{Bayesian} & \multirow{3}{*}{100} & $H$ & 1,500 & 1,513 & 120 & 0.01 & 0.08 & 0.94 & 466\\
        & & $\ps$ & 0.7 & 0.70 & 0.05 & -0.00 & 0.07 & 0.94 & 0.20\\
        & & $\pmay$ & 0.2 & 0.20 & 0.04 & 0.02 & 0.20 & 0.94 & 0.15\\        
        \hline
    \end{tabular}
    }
\end{table}

Moving on to model $\MI$, we present the results of this analysis in Table~\ref{tab:mod2}, which demonstrate a pattern akin to that observed in model $\MB$. The Bayesian estimator of $H$ yields a larger bias, while the MLE has a coverage probability farther from 95\%. And the gap diminishes when transitioning to the large city scenario. Besides, the RRMSE and LCI follow a similar pattern as observed in model $\MB$.

\begin{table}
    \centering
    \caption{Results of the simulation studies with $\MI$ for the MLEs and the Bayesian estimators. All the values are rounded to integers or 2 decimal points.}
    \label{tab:mod2}
    \resizebox{\textwidth}{!}{
    \begin{tabular}{ccccccccccc}
        \hline
        Method & $M$ & Parameter & True Value & Estimate & SD & RBias & RRMSE & CP & LCI\\
        \hline
        \multirow{4}{*}{MLE} & \multirow{4}{*}{15} & $H$ & 150 & 150 & 29 & 0.00 & 0.22 & 0.88 & 118\\
        & & $\ps$ & 0.7 & 0.72 & 0.12 & 0.03 & 0.18 & 0.98 & 0.46\\
        & & $\pmayin$ & 0.2 & 0.21 & 0.13 & 0.04 & 0.83 & 0.96 & 0.70\\
        & & $\pins$ & 0.8 & 0.80 & 0.04 & -0.00 & 0.05 & 0.95 & 0.15\\
        \hline
         \multirow{4}{*}{Bayesian} & \multirow{4}{*}{15} & $H$ & 150 & 159 & 38 & 0.06 & 0.22 & 0.97 & 142\\
        & & $\ps$ & 0.7 & 0.68 & 0.11 & -0.03 & 0.16 & 0.97 & 0.43\\
        & & $\pmayin$ & 0.2 & 0.26 & 0.14 & 0.30 & 0.73 & 0.96 & 0.52\\
        & & $\pins$ & 0.8 & 0.80 & 0.04 & -0.01 & 0.05 & 0.95 & 0.14\\
        \hline
        \multirow{4}{*}{MLE} & \multirow{4}{*}{100} & $H$ & 1,500 & 1,498 & 107 & -0.00 & 0.07 & 0.93 & 420\\
        & & $\ps$ & 0.7 & 0.70 & 0.05 & 0.01 & 0.07 & 0.94 & 0.19\\
        & & $\pmayin$ & 0.2 & 0.20 & 0.06 & -0.00 & 0.30 & 0.97 & 0.24\\
        & & $\pins$ & 0.8 & 0.80 & 0.01 & 0.00 & 0.02 & 0.96 & 0.05\\
        \hline
         \multirow{4}{*}{Bayesian} & \multirow{4}{*}{100} & $H$ & 1,500 & 1,512 & 111 & 0.01 & 0.07 & 0.94 & 433\\
        & & $\ps$ & 0.7 & 0.70 & 0.05 & -0.00 & 0.07 & 0.94 & 0.18\\
        & & $\pmayin$ & 0.2 & 0.21 & 0.06 & 0.04 & 0.29 & 0.96 & 0.23\\
        & & $\pins$ & 0.8 & 0.80 & 0.01 & 0.00 & 0.02 & 0.96 & 0.05\\        
        \hline
    \end{tabular}
    }
\end{table}

Finally, in the context of model $\MC$, the analysis results presented in Table~\ref{tab:mod3} offer insights into the performance of this model. Our findings in $\MC$ echo the trends identified in previous simulation studies, particularly for the large city scenario. However, in the small city scenario, we observe some deviations from the trends observed earlier. Specifically, while the coverage probability of the MLE of $H$ is the same as that of the Bayesian estimator, the MLE has a smaller relative bias (4\% compared to 9\%) with a larger RRMSE and a larger LCI. The difference could be attributed to the small sample size ($M$), resulting in less information available in each class, especially within the classes with small capture probabilities. Further discussion will be provided in Section~\ref{discussion}.

\begin{table}
    \centering
    \caption{Results of the simulation studies with $\MC$ for the MLEs and the Bayesian estimators. All the values are rounded to integers or 2 decimal points.}
    \label{tab:mod3}
    \resizebox{\textwidth}{!}{
    \begin{tabular}{ccccccccccc}
        \hline
        Method & $M$ & Parameter & True Value & Estimate & SD & RBias & RRMSE & CP & LCI\\
        \hline
        \multirow{5}{*}{MLE} & \multirow{5}{*}{30} & $H$ & 300 & 313 & 65 & 0.04 & 0.25 & 0.97 & 358\\
        & & $\ps_1$ (easy) & 0.9 & 0.90 & 0.07 & 0.00 & 0.08 & 0.96 & 0.37\\
        & & $\ps_2$ (hard) & 0.4 & 0.42 & 0.14 & 0.06 & 0.38 & 0.97 & 0.52\\
        & & $\pmayin$ & 0.2 & 0.19 & 0.10 & -0.07 & 0.54 & 0.98 & 0.46\\
        & & $\pins$ & 0.8 & 0.80 & 0.03 & 0.00 & 0.03 & 0.95 & 0.10\\
        \hline
        \multirow{5}{*}{Bayesian} & \multirow{5}{*}{30} & $H$ & 300 & 326 & 87 & 0.09 & 0.20 & 0.97 & 314\\
        & & $\ps_1$ (easy) & 0.9 & 0.84 & 0.08 & -0.06 & 0.09 & 0.94 & 0.32\\
        & & $\ps_2$ (hard) & 0.4 & 0.42 & 0.13 & 0.04 & 0.32 & 0.96 & 0.49\\
        & & $\pmayin$ & 0.2 & 0.22 & 0.10 & 0.09 & 0.49 & 0.97 & 0.38\\
        & & $\pins$ & 0.8 & 0.80 & 0.03 & -0.00 & 0.03 & 0.94 & 0.10\\
        \hline
        \multirow{5}{*}{MLE} & \multirow{5}{*}{100} & $H$ & 1,500 & 1,510 & 142 & 0.01 & 0.10 & 0.97 & 702\\
        & & $\ps_1$ (easy) & 0.9 & 0.91 & 0.04 & 0.01 & 0.05 & 0.93 & 0.16\\
        & & $\ps_2$ (hard) & 0.4 & 0.40 & 0.08 & 0.01 & 0.20 & 0.97 & 0.30\\
        & & $\pmayin$ & 0.2 & 0.20 & 0.06 & -0.01 & 0.30 & 0.96 & 0.23\\
        & & $\pins$ & 0.8 & 0.80 & 0.01 & -0.00 & 0.02 & 0.95 & 0.05\\
        \hline
        \multirow{5}{*}{Bayesian} & \multirow{5}{*}{100} & $H$ & 1,500 & 1,535 & 155 & 0.02 & 0.10 & 0.96 & 601\\
        & & $\ps_1$ (easy) & 0.9 & 0.89 & 0.04 & -0.01 & 0.05 & 0.96 & 0.16\\
        & & $\ps_2$ (hard) & 0.4 & 0.40 & 0.08 & 0.01 & 0.19 & 0.96 & 0.30\\
        & & $\pmayin$ & 0.2 & 0.21 & 0.06 & 0.04 & 0.29 & 0.95 & 0.23\\
        & & $\pins$ & 0.8 & 0.80 & 0.01 & -0.00 & 0.02 & 0.95 & 0.05\\        
        \hline
    \end{tabular}
    }
\end{table}

Overall, both MLE and Bayesian estimators provide an accurate estimate of the target population size. The findings from the simulation studies shed light on the trade-off between bias and coverage probability in our models. The choice between MLE and Bayesian methods should be made based on the specific characteristics of the models and the desired objectives of the analysis.

\section{Application to the S-Night Street Enumeration Survey} \label{application}

In this section, we apply our method to the 1990 S-Night data. On the night of March 20-21, 1990, the United States Census Bureau carried out the Shelter and Street Night Enumeration, also known as S-Night \citep{barrett19921990}. The survey was conducted in five major cities: New Orleans, New York, Phoenix, Los Angeles and Chicago. Prior to the enumeration, a known number of plants, trained to dress and act like homeless people, were deployed at designated sites. The plants were instructed to stay in an open area to allow the enumerators to see and enumerate them during street enumeration between 2 to 4 a.m., and enumerators were asked to interview all individuals encountered in the pre-assigned sites. After the enumeration, the plants were requested to fill out questionnaires to report whether an enumerator interviewed them and whether they believed they were counted by an enumerator if not interviewed. For more details refer to \citet{martin1992assessment}. Note that this survey is designed to estimate the homeless population size present in the areas targeted during the survey; it is not meant to provide an exhaustive count of the homeless population in the entire city.

The data we use for our demonstration are shown in Table~\ref{tab:data}. Given the time elapsed since the original study, the original data could not be retrieved, therefore we reconstructed Table~\ref{tab:data} approximately from data summaries published in \citet{martin1992assessment} and \citet{martin1997issues}. Notably, while the number of plants interviewed could be reconstructed from the literature, the number of homeless interviewed and the exact counts in easy/hard sites were not available. Thus our methodology is demonstrated using a modified version of model $\MI$. In this variant, Equation~\eqref{eq:IIHi} is omitted due to the unavailability of $\Hi$.

\begin{table}
    \centering
    \caption{The 1990 S-Night data reconstructed from the literature.}
    \label{tab:data}
    \begin{tabular}{lrrrrr}
        \hline
        & \multicolumn{1}{l}{} & \multicolumn{1}{l}{New} & \multicolumn{1}{l}{} & \multicolumn{1}{l}{New} & \multicolumn{1}{l}{Los} \\
        & \multicolumn{1}{l}{Chicago} & \multicolumn{1}{l}{Orleans} & \multicolumn{1}{l}{Phoenix} & \multicolumn{1}{l}{York} & \multicolumn{1}{l}{Angeles} \\
        \hline
        Plants ($M$) & 13 & 58 & 26 & 94 & 25 \\
        \quad Interviewed ($M^i$) & 2 & 41 & 18 & 40 & 16 \\
        \quad Yes ($M^y$)   & 0 & 6 & 3 & 19 & 1 \\
        \quad Maybe ($M^m$) & 5 & 5 & 1 & 13 & 2 \\
        \quad No ($M^n$)    & 6 & 6 & 4  & 22 & 6 \\
        Census ($Y$) & 11 & 109 & 104 & 1,240 & 217 \\
        \hline
    \end{tabular}
\end{table}

We conduct the data analysis separately for each city. The results for both MLEs and Bayesian estimators (posterior medians) are presented in Table~\ref{tab:app}. For the Bayesian inference, we applied the prior specifications described in Section~\ref{sim}. For our target, the homeless population size $H$, the Bayesian method provided larger estimates and standard deviations compared to MLE, except for Chicago. This could be due to a sensitivity to the prior, which tends to inflate estimates slightly. However, Chicago stands out with significantly distinct results between Bayesian estimators and MLEs in comparison to other cities. This discrepancy arises from the absence of plants self-assessing as ``yes'' in Chicago. As a result, the MLE of $\pins$ is 1 (which is an overestimate) with an estimated variance of zero. In contrast, Bayesian estimates are influenced by the prior setting, introducing variability and leading to divergent results between the two methods.

Although we do not have the true value for these parameters, the consistency in estimation outcomes suggests that both Bayesian and MLE approaches offer valuable insights. The choice between the methods should be guided by the specific analytical needs and the availability of prior information, ensuring researchers can harness the most suitable technique for their particular context.

\begin{table}
    \centering
    \caption{Results of the application to the S-Night data using $\MI$ without Equation~\eqref{eq:IIHi} separately in each city. All the values are rounded to integers or 2 decimal points.}
    \label{tab:app}
    \begin{tabular}{ccccccc}
    \hline
     & \multicolumn{3}{c}{Bayesian} & \multicolumn{3}{c}{MLE}\\
    \hline
       Parameter & Estimate & SD & 95\% CrI & Estimate & SD & 95\% CI\\
    \hline
       \multicolumn{7}{c}{Chicago}\\
    \hline
        $H$ & 37 & 40 & (11, 156) & 54 & 38 & (13, 217)\\
        $\ps$ & 0.22 & 0.12 & (0.06, 0.51) & 0.16 & 0.10 & (0.04, 0.46)\\
        $\pmayin$ & 0.46 & 0.13 & (0.21, 0.72) & 0.45 & 0.15 & (0.20, 0.73)\\
        $\pins$ & 0.71 & 0.22 & (0.21, 0.99) & 1.00 & 0.00 & (1.00, 1.00)\\
    \hline
     \multicolumn{7}{c}{New Orleans}\\
    \hline
        $H$ & 70 & 7 & (61, 87) & 69 & 6 & (58, 82)\\
        $\ps$ & 0.84 & 0.05 & (0.73, 0.93) & 0.86 & 0.05 & (0.73, 0.94)\\
        $\pmayin$ & 0.31 & 0.10 & (0.13, 0.54) & 0.29 & 0.11 & (0.13, 0.54)\\
        $\pins$ & 0.82 & 0.06 & (0.69, 0.91) & 0.83 & 0.06 & (0.68, 0.91)\\
    \hline
     \multicolumn{7}{c}{Phoenix}\\
    \hline
        $H$ & 102 & 12 & (87, 135) & 98 & 10 & (80, 120)\\
        $\ps$ & 0.81 & 0.08 & (0.64, 0.93) & 0.84 & 0.08 & (0.64, 0.94)\\
        $\pmayin$ & 0.18 & 0.12 & (0.03, 0.49) & 0.12 & 0.12 & (0.02, 0.54)\\
        $\pins$ & 0.82 & 0.08 & (0.63, 0.94) & 0.84 & 0.08 & (0.61, 0.94)\\
    \hline
     \multicolumn{7}{c}{New York}\\
    \hline
        $H$ & 1,709 & 142 & (1,494, 2,056) & 1,688 & 131 & (1,450, 1,964)\\
        $\ps$ & 0.69 & 0.05 & (0.57, 0.78) & 0.70 & 0.05 & (0.59, 0.79)\\
        $\pmayin$ & 0.25 & 0.06 & (0.15, 0.37) & 0.24 & 0.06 & (0.14, 0.37)\\
        $\pins$ & 0.61 & 0.06 & (0.49, 0.73) & 0.61 & 0.06 & (0.48, 0.73)\\
    \hline
     \multicolumn{7}{c}{Los Angeles}\\
    \hline
        $H$ & 290 & 47 & (233, 415) & 282 & 40 & (215, 372)\\
        $\ps$ & 0.69 & 0.09 & (0.49, 0.84) & 0.71 & 0.09 & (0.50, 0.86)\\
        $\pmayin$ & 0.26 & 0.13 & (0.07, 0.56) & 0.22 & 0.14 & (0.06, 0.58)\\
        $\pins$ & 0.89 & 0.08 & (0.69, 0.98) & 0.92 & 0.07 & (0.63, 0.99)\\
    \hline
    \end{tabular}
\end{table}

As a point of comparison, Supplementary Table C4 presents the estimates and the 95\% confidence intervals obtained using the hybrid Chapman-Bailey estimator, as described in \citet{laska1993plant}. This estimator is calculated under two extreme scenarios considered in \citet{martin1997issues}, representing contrasting treatments of ``maybe'' responses. In one scenario all the plants who self-assessed as ``maybe'' are considered as ``yes'', while in the other, they are treated as ``no'', since the Chapman-Bailey estimator cannot handle uncertain assessments. Our estimates are all included in the 95\% CI for both scenarios, except for Chicago, where only the 95\% CI under the second scenario includes our estimates.

\section{Discussion} \label{discussion}

In this work, we have introduced a novel plant-capture modeling framework that incorporates uncertain assessment of capture and can allow for partial identification of plants as well as heterogeneity across survey sites. Within this framework, we have proposed two distinct inference approaches: frequentist maximum likelihood estimation and a Bayesian methodology. Our simulation studies have demonstrated the Bayesian approach's ability to achieve coverage probabilities close to the desired 95\% while exhibiting a slightly larger bias compared to the MLEs. In contrast, MLEs tend to have a smaller bias but the coverage probabilities can occasionally deviate from the ideal 95\% in small population settings. Importantly, our simulations have revealed that as the population size increases, the discrepancy between the two methods diminishes, and their performance improves simultaneously. However, an exception arises under $\MC$: it may have suboptimal performance due to insufficient information about different sites, especially when the population size in each site is relatively small. Furthermore, we have applied these models to estimate the homeless population size using the 1990 S-Night data, shedding light on their real-world applicability. 

The insights gained from this research have the potential to significantly contribute to public health planning and policy formulation, especially with regard to addressing the needs of vulnerable populations. Estimating the vulnerable population size within a society holds immense importance due to the multifaceted impact on their quality of life. Access to housing play vital roles in education and labor market participation. Moreover, quantifying vulnerable populations, such as the homeless population, is instrumental in shaping governmental policies, particularly those related to housing provisions, and enables the evaluation of intervention effectiveness \citep{coumans2017estimating}. 

To explore the applicability of our models in real-world scenarios, further investigations are warranted, with a particular focus on assessing the validity of the independence assumptions. The independence (MAR) assumptions introduced in Section~\ref{model} play a crucial role in our models, allowing estimation of the size of the target population despite uncertainty arising from the plants self-assessed as ``maybe''. However, it is important to recognize that these assumptions are essentially untestable and may not hold under certain circumstances, particularly when there is a preference for ``yes'' or ``no'' responses among the ``maybe'' category. An example in the homeless survey could be, if a significant portion of ``maybe'' plants were sleeping in locations that were difficult for enumerators to observe, these plants would be more likely to go unnoticed, thus violating the assumption of independence, unless sleeping locations are used to define classes within model $\MC$. However, there is a limit to the number of classes that should be used as the model can fail to provide accurate estimators in the cases when either no plants self-assess as ``yes'', or no plants self-assess as ``no''. Under these situations, all the plants certain about their status belong to a single group, leaving no information available about the other group. As a result, all the plants self-assessing as ``maybe'' will be categorized into the same group, which is an extreme ratio for the ``maybe'' plants and potentially leads to an imprecise estimator, especially when the number of ``maybe'' plants is relatively large. For this reason, we recommend that $\MC$ presented in Section~\ref{model3} should only be applied when there are sufficient plants in each class to avoid this issue. Or, one could pool information about unknown parameters across sites with the help of a random-effects model. Another potential avenue for improvement is to consider a more flexible modeling approach. Instead of setting equality in Assumptions~\ref{assumpI3},~\ref{assumpII3} and~\ref{assumpIII3}, it may be possible to relax the MAR assumptions by treating these probabilities as exchangeable via a hierarchical prior that controls the degree to which the MAR assumption is relaxed. This would propagate additional uncertainty into the final estimates. 

Finally, there may be room for improvement in addressing variations between different survey sites through model refinements. For instance, a more sophisticated approach might involve modeling site-specific probabilities using logistic regression by incorporating covariates such as site characteristics and GPS location data. Notably, the Counting Us Mobile App \citep{CountUs} has already demonstrated its capability to track the locations of enumerators and plants during surveys, facilitating point-in-time counts in 50 regions across the United States. Utilizing data on the distances between plants and enumerators could potentially lead to a more accurate estimation of capture probabilities.  This could enable adjustments in cases where enumerators were delayed, positioned inaccurately, or absent altogether. Additionally, such detailed spatial data could provide valuable insights into the validity of our MAR assumptions, allowing for the modeling of deviations from this assumption and enhancing the robustness of population size estimates.

\section*{Acknowledgements}

The authors thank Professor Ruth King for helpful discussions. This research is supported by NSERC Discovery Grants RGPIN-2020-04364 (ML), RGPIN-2019-04404 (AB), and the CANSSI Collaborative Research Team Grant on Modern Techniques for Survey Sampling and Complex Data (YW, AB).\vspace*{-8pt}

\bibliographystyle{chicago}
\bibliography{ref} 

\begin{appendices}

\setcounter{table}{0}
\renewcommand{\thetable}{\Alph{section}\arabic{table}}

\section{Derivation of MLE for Model $\MB$}

Based on the joint likelihood of the parameters of interest $\bm{\gamma}$ described in Section~\ref{mle}, it is easy to get the log-likelihood as
\begin{align}
    \nonumber l(\bm{\gamma}&; y, m^{yes}, m^{mb}, m^{no})\\
    \nonumber =\,& m^{yes}\log\{\ps(1-\pmay)\} + m^{mb}\log\pmay + m^{no}\log\{(1-\ps)(1-\pmay)\}\\
    \nonumber &+ \log\{(H+m^{mb})!\} - \log\{(H+m^{mb}-y+m^{yes})!\}\\
    &+ (y-m^{yes})\log\ps + (H+m^{mb}-y+m^{yes})\log(1-\ps) \tag{A.1}\label{eq:loglik}
\end{align}

Taking partial derivatives on Equation~\eqref{eq:loglik} with respect to $\pmay$ and $\ps$ and letting them equal to 0, we arrive at the ML estimator for $\pmay$ as $\frac{\Mm}{\My+\Mm+\Mn}=\frac{\Mm}{M}$ and the ML estimator for $\ps$ (with known $H$) as $\frac{Y}{H+\My+\Mm+\Mn} = \frac{Y}{H+M}$. 

To find the ML estimator for $H$, we consider the ratio
\begin{align*}
    \frac{L(H+1)}{L(H)} &= \frac{\left( \begin{array}{c} H+1+m^{mb} \\ y-m^{yes} \end{array} \right)(1-\ps)^{H+1+m^{mb}-y+m^{yes}}}{\left( \begin{array}{c} H+m^{mb} \\ y-m^{yes} \end{array} \right)(1-\ps)^{H+m^{mb}-y+m^{yes}}}\\
    &=\frac{H+1+m^{mb}}{H+1+m^{mb}-y+m^{yes}}(1-\ps) < 1 \text{ when } H>\frac{y-m^{yes}-m^{mb}\ps}{\ps}-1.
\end{align*}
This implies a ML estimator of $H$ (when $\ps$ is known) as $\lfloor \frac{Y-\My-\Mm\ps}{\ps}\rfloor$. The ML estimator for $\pmay$ does not depend on the other two parameters, but the ML estimators for $H$ and $\ps$ depend on each other. Since the ML estimators need to satisfy all three expressions simultaneously, we can solve for $H$ and $\ps$ to yield the ML estimators for each parameter: $\hat{\ps} = \frac{\My}{\My+\Mn}$, $\hat{\pmay} = \frac{\Mm}{M}$ and $\hat{H} = \lfloor \frac{Y-\My-\Mm\hat{\ps}}{\hat{\ps}} \rfloor$, where $\lfloor x\rfloor$ denotes the greatest integer less than or equal to $x$. 

\section{Alternative Computational Methods for Bayesian Inference}

We explored two alternative computational approaches to fit our models. These approaches are detailed in this section, and evaluated in a simulation study in Appendix~\ref{webtable}.

\subsection{Bayesian Normal Approximation (BNA)}\label{BNA}

Bayesian normal approximation is an approach that constructs an approximate representation of the posterior distribution using a multivariate normal distribution. The mean of the distribution is approximated by the vector of posterior mode of the parameters, obtained via numerical optimization of the posterior distribution. The variance-covariance matrix of the distribution is defined as the inverse of the negative Hessian matrix of the log posterior density at the modes. A more comprehensive description of this technique is given in \citet{bda}. 

Similarly to the MLE method described in Section~\ref{mle}, we can express the posterior distribution of our proposed models in a general form $\pi(\bm{\gamma|\bm{x}}) \propto \pi(\bm{\gamma}) P(\bm{X}=\bm{x}|\bm{\gamma}) = \pi(\bm{\gamma})\sum_{\bm{z} \in \Omega} P(\bm{X}=\bm{x},\bm{Z}=\bm{z}|\bm{\gamma})$, where $\bm{\gamma}$ denotes the model parameters with a prior distribution $\pi(\bm{\gamma})$, $\bm{X}$ represents the data, $\bm{Z}$ stands for latent variables, and $\Omega$ denotes the set of values that $\bm{Z}$ can take. We also apply a log transformation on the counts and a logit transformation on the probabilities to remove any constraints on their bounds, avoiding computational issues related to boundary constraints. The numerical method to apply this approach is the same as the MLE approach detailed in Section~\ref{mle}. Furthermore, the prior settings for the parameters remain the same with those described in Section~\ref{sim}.

\subsection{Uncertainty Propagation Method (UP)}\label{UP}

While inference for model $\MB$ is relatively straightforward using MCMC algorithms, the specification of models $\MI$ and $\MC$ using probabilistic programming languages can be relatively complicated because of the equality constraints in Equations~\eqref{eq:IIY} and \eqref{eq:IIIY}, as discussed in Section~\ref{bayes}. Instead of using the \textit{dsum} function in JAGS, another simple solution is to employ an uncertainty propagation (UP) method to obtain an approximate posterior inference for $H$. To illustrate this method, we use model $\MI$ as an example.

Initially, it is important to recognize that if $\ps$ and $\Hs$ were observed, it would be feasible to construct an approximate representation of the posterior distribution $\pi(H | \ps, \Hs)$ using a normal distribution due to the Bernstein–von Mises theorem, as follows:
\begin{equation}
    H | \Hs, \ps \sim N\left( \Ho, \frac{\Ho(1-\ps)}{{\ps}}\right),\tag{B.2.1}\label{eq:norm}
\end{equation}
where $\Ho = \Hs/\ps$ is a variant of the MLE for $H$ \citep{rukhin1975statistical} based on the binomial distribution in Equation~\eqref{eq:IIH}. A proof of the asymptotic correspondence of $\Ho$ with the MLE is presented in the box below.

\begin{tcolorbox}[colback=red!5!white,title=Proof]
Suppose we have $\Hs\sim\dbin(H, \ps)$. Given $\ps$ and $\Hs$, we use the method introduced in the Appendix of the main paper to derive the MLE of $H$:

\begin{align*}
    \frac{L(H+1)}{L(H)} &= \frac{\frac{(H+1)!}{\Hs!(H+1-\Hs)!}(\ps)^{\Hs}(1-\ps)^{H+1-\Hs}}{\frac{H!}{\Hs!(H-\Hs)!}(\ps)^{\Hs}(1-\ps)^{H-\Hs}}\\
    &=\frac{H+1}{H+1-\Hs}(1-\ps) < 1 \text{ when } H>\frac{\Hs}{\ps}-1,
\end{align*}
which leads to the ML estimator of $H$ as $\lfloor \frac{\Hs}{\ps}\rfloor$.
\end{tcolorbox}

\noindent The approximate MLE, $\Ho$, has variance
\begin{equation*}
    \Var(\Ho) = \frac{\Var(\Hs)}{{(\ps)}^2} = \frac{H(1-\ps)}{\ps},
\end{equation*}
which is estimated by $\frac{\Ho(1-\ps)}{{\ps}}$ in Equation~\eqref{eq:norm}. Hence, given observed values of $\ps$ and $\Hs$, we could sample from the approximate posterior distribution by sampling directly from Equation~\eqref{eq:norm}.

In practice, while $\ps$ and $\Hs$ are not directly observable, we can sample from their posterior distribution conditional on the observed data $\bm{x}$, $\pi(\Hs, \ps | \bm{x})$. This is possible because $\ps$ and $\Hs$ are completely informed when Equation~\eqref{eq:IIH} is removed from our model. In fact, the role of Equation~\eqref{eq:IIH} is strictly to expand $\Hs$ into $H$ via $\ps$, offering no insights into any model parameters other than $H$. To summarize, fitting model $\MI$ without Equation~\eqref{eq:IIH} and marginalizing over $\pins, \pmay$ and $\Mms$ provides a posterior sample from $L(\Hs, \ps | \bm{x})$. Once an MCMC sample is obtained (first step), the values can be plugged into Equation~\eqref{eq:norm} to simulate posterior samples from $H$ (second step), which result in the desired approximate posterior distribution $\pi(H| \bm{x})$. Essentially, uncertainty from the first step is propagated into the second step.

Instead of employing a two-step approach, the desired outcome can be achieved more straightforwardly in a single step by replacing Equation~\eqref{eq:IIH} in our model with the approximate normal distribution~\eqref{eq:norm}. This substitution simplifies the MCMC process and enhances computational efficiency. BUGS-based software can thus be used to directly sample from our approximate representation of $\pi(\bm{\gamma}| \bm{x})$, which can then be marginalized to sample from $\pi(H| \bm{x})$.

Finally, when implementing the proposed UP method in BUGS languages, the following grammatical nuance must be considered: Equations~\eqref{eq:IIY} and \eqref{eq:IIIY} need to be rearranged to shift $\Hs$ to the left side. This ensures that the software doesn't interpret $\Hs$ as an undefined node.

\section{Supplementary Tables}\label{webtable}

\begin{table}[!htbp]
    \centering
    \caption{Results of the simulation studies for $\MB$ using BNA. All the values are rounded to integers or 2 decimal points.}
    \label{tab:supmod1}
    \resizebox{\textwidth}{!}{
    \begin{tabular}{ccccccccccc}
        \hline
        Method & $M$ & Parameter & True Value & Median & SD & RBias & RRMSE & CP & LCI\\
        \hline
        \multirow{3}{*}{BNA} & \multirow{3}{*}{15} & $H$ & 150 & 149 & 27 & -0.01 & 0.18 & 0.91 & 109\\
        & & $\ps$ & 0.7 & 0.72 & 0.11 & 0.03 & 0.15 & 0.99 & 0.43\\
        & & $\pmay$ & 0.2 & 0.24 & 0.10 & 0.19 & 0.49 & 0.94 & 0.39\\
        \hline
        \multirow{3}{*}{BNA} & \multirow{3}{*}{100} & $H$ & 1,500 & 1,498 & 112 & 0.00 & 0.08 & 0.94 & 442\\
        & & $\ps$ & 0.7 & 0.70 & 0.05 & 0.01 & 0.07 & 0.95 & 0.19\\
        & & $\pmay$ & 0.2 & 0.21 & 0.04 & 0.03 & 0.20 & 0.96 & 0.16\\
        \hline
    \end{tabular}
    }
\end{table}

\begin{table}
    \centering
    \caption{Results of the simulation studies for $\MI$ using BNA and UP method. All the values are rounded to integers or 2 decimal points.}
    \label{tab:supmod2}
    \resizebox{\textwidth}{!}{
    \begin{tabular}{cccccccccc}
        \hline
        Method & $M$ & Parameter & True Value & Median & SD & RBias & RRMSE & CP & LCI\\
        \hline
        \multirow{4}{*}{BNA} & \multirow{4}{*}{15} & $H$ & 150 & 150 & 26 & 0.00 & 0.17 & 0.92 & 105\\
        & & $\ps$ & 0.7 & 0.71 & 0.11 & 0.02 & 0.14 & 0.98 & 0.42\\
        & & $\pmayin$ & 0.2 & 0.28 & 0.15 & 0.39 & 0.73 & 0.93 & 0.55\\
        & & $\pins$ & 0.8 & 0.79 & 0.04 & -0.01 & 0.05 & 0.94 & 0.15\\
        \hline
        \multirow{5}{*}{UP} & \multirow{5}{*}{15} & $\hat{H}$ & 150 & 159 & 38 & 0.06 & 0.22 & 0.97 & 144\\
        & & $\hat{H}_0$ & 150 & 159 & 37 & 0.06 & 0.22 & 0.96 & 138\\
        & & $\ps$ & 0.7 & 0.68 & 0.11 & -0.03 & 0.16 & 0.97 & 0.43\\
        & & $\pmayin$ & 0.2 & 0.26 & 0.14 & 0.30 & 0.73 & 0.96 & 0.52\\
        & & $\pins$ & 0.8 & 0.80 & 0.04 & -0.01 & 0.05 & 0.95 & 0.14\\
        \hline
        \multirow{4}{*}{BNA} & \multirow{4}{*}{100} & $H$ & 1,500 & 1,500 & 105 & 0.00 & 0.07 & 0.93 & 414\\
        & & $\ps$ & 0.7 & 0.70 & 0.05 & 0.00 & 0.07 & 0.94 & 0.18\\
        & & $\pmayin$ & 0.2 & 0.21 & 0.06 & 0.06 & 0.29 & 0.96 & 0.23\\
        & & $\pins$ & 0.8 & 0.80 & 0.01 & 0.00 & 0.02 & 0.95 & 0.05\\
        \hline
        \multirow{5}{*}{UP} & \multirow{5}{*}{100} & $\hat{H}$ & 1,500 & 1,512 & 111 & 0.01 & 0.07 & 0.94 & 434\\
        & & $\hat{H}_0$ & 1,500 & 1,513 & 108 & 0.01 & 0.07 & 0.93 & 422\\
        & & $\ps$ & 0.7 & 0.70 & 0.05 & -0.00 & 0.07 & 0.94 & 0.18\\
        & & $\pmayin$ & 0.2 & 0.21 & 0.06 & 0.04 & 0.29 & 0.96 & 0.23\\
        & & $\pins$ & 0.8 & 0.80 & 0.01 & 0.00 & 0.02 & 0.96 & 0.05\\
        \hline
    \end{tabular}
    }
\end{table}

\begin{table}
    \centering
    \caption{Results of the simulation studies for $\MC$ using BNA and UP method. All the values are rounded to integers or 2 decimal points.}
    \label{tab:supmod3}
    \resizebox{\textwidth}{!}{
    \begin{tabular}{cccccccccc}
        \hline
        Method & $M$ & Parameter & True Value & Median & SD & RBias & RRMSE & CP & LCI\\
        \hline
        \multirow{5}{*}{BNA} & \multirow{5}{*}{30} & $H$ & 300 & 294 & 40 & -0.02 & 0.12 & 0.99 & 218\\
        & & $\ps_{easy}$ & 0.9 & 0.86 & 0.08 & -0.04 & 0.08 & 0.94 & 0.33\\
        & & $\ps_{hard}$ & 0.4 & 0.48 & 0.13 & 0.19 & 0.35 & 0.95 & 0.48\\
        & & $\pmayin$ & 0.2 & 0.23 & 0.10 & 0.14 & 0.49 & 0.95 & 0.41\\
        & & $\pins$ & 0.8 & 0.80 & 0.03 & -0.00 & 0.03 & 0.96 & 0.10\\
        \hline
        \multirow{6}{*}{UP} & \multirow{6}{*}{30} & $\hat{H}$ & 300 & 327 & 103 & 0.09 & 0.21 & 0.96 & 323\\
        & & $\hat{H}_0$ & 300 & 328 & 100 & 0.09 & 0.21 & 0.95 & 312\\
        & & $\ps_{easy}$ & 0.9 & 0.84 & 0.08 & -0.06 & 0.09 & 0.94 & 0.32\\
        & & $\ps_{hard}$ & 0.4 & 0.42 & 0.13 & 0.04 & 0.32 & 0.96 & 0.49\\
        & & $\pmayin$ & 0.2 & 0.22 & 0.10 & 0.09 & 0.49 & 0.98 & 0.38\\
        & & $\pins$ & 0.8 & 0.80 & 0.03 & -0.00 & 0.03 & 0.94 & 0.10\\
        \hline
        \multirow{5}{*}{BNA} & \multirow{5}{*}{100} & $H$ & 1,500 & 1,486 & 124 & -0.01 & 0.08 & 0.98 & 634\\
        & & $\ps_{easy}$ & 0.9 & 0.90 & 0.04 & -0.00 & 0.05 & 0.96 & 0.17\\
        & & $\ps_{hard}$ & 0.4 & 0.42 & 0.08 & 0.06 & 0.20 & 0.95 & 0.29\\
        & & $\pmayin$ & 0.2 & 0.21 & 0.06 & 0.06 & 0.29 & 0.96 & 0.23\\
        & & $\pins$ & 0.8 & 0.80 & 0.01 & -0.00 & 0.02 & 0.95 & 0.05\\
        \hline
        \multirow{6}{*}{UP} & \multirow{6}{*}{100} & $\hat{H}$ & 1,500 & 1,538 & 156 & 0.03 & 0.10 & 0.96 & 606\\
        & & $\hat{H}_0$ & 1,500 & 1,538 & 152 & 0.03 & 0.10 & 0.95 & 589\\
        & & $\ps_{easy}$ & 0.9 & 0.89 & 0.04 & -0.01 & 0.05 & 0.96 & 0.16\\
        & & $\ps_{hard}$ & 0.4 & 0.41 & 0.08 & 0.00 & 0.19 & 0.96 & 0.30\\
        & & $\pmayin$ & 0.2 & 0.21 & 0.06 & 0.04 & 0.29 & 0.95 & 0.23\\
        & & $\pins$ & 0.8 & 0.80 & 0.01 & -0.00 & 0.02 & 0.95 & 0.05\\
        \hline
    \end{tabular}
    }
\end{table}

\begin{table}
    \centering
    \caption{Estimation of the homeless population size $H$ using the Chapman-Bailey estimator. All the values are rounded to integers.}
    \label{tab:laska}
    \begin{tabular}{ccccc}
    \hline
    & \multicolumn{2}{c}{Maybe as seen} & \multicolumn{2}{c}{Maybe as not seen} \\
    \hline
    City & Estimate & 95\% CI & Estimate & 95\% CI\\
    \hline
    Chicago & 7 & (3, 20) & 42 & (23, $\infty$) \\
    New Orleans & 63 & (56, 72) & 76 & (65, 91) \\
    Phoenix & 96 & (81, 118) & 102 & (85, 129)\\
    New York & 1,520 & (1,368, 1,721) & 1,670 & (1,624, 2,233)\\
    Los Angeles & 257 & (212, 335) & 289 & (231,402)\\
    \hline
    \end{tabular}
\end{table}

\end{appendices}

\end{document}